\documentclass[12pt]{iopart}


\expandafter\let\csname equation*\endcsname\relax
\expandafter\let\csname endequation*\endcsname\relax
\usepackage{amsmath}
\usepackage{cite}

\usepackage{hyperref}
\usepackage{amsfonts}
\usepackage{bbm}
\usepackage{graphicx} 
\usepackage{verbatim}
\usepackage[normalem]{ulem}
\usepackage{algorithm}
\usepackage{tabularx}
\usepackage{url} 
\usepackage{empheq}
\usepackage{adjustbox}
\usepackage[T1]{fontenc}
\usepackage[utf8]{inputenc}
\usepackage{algpseudocode}
\usepackage{appendix}

\begin{document}


\title{Amortized Simulation-Based Frequentist Inference for Tractable and Intractable Likelihoods}


\author{Ali Al Kadhim}
\address{Department of Physics, Florida State University, Tallahassee, FL 32306-4350}
\ead{aa18dg@fsu.edu}

\author{Harrison B. Prosper}
\address{Department of Physics, Florida State University, Tallahassee, FL 32306-4350}

\author{Olivia F. Prosper}
\address{Department of Mathematics, University of Tennessee, Knoxville, TN 37996-1320}



\begin{abstract}
High-fidelity simulators that connect theoretical models with observations are indispensable tools in many sciences. When coupled with machine learning, a simulator makes it possible to infer the parameters of a theoretical model directly from real and simulated observations without explicit use of the likelihood function. This is of particular interest when the latter is intractable. In this work, we introduce a simple extension of the recently proposed likelihood-free frequentist inference (\verb|LF2I|) approach that has some computational advantages. Like \verb|LF2I|, this extension yields provably valid confidence sets in parameter inference problems in which a high-fidelity simulator is available. The utility of our algorithm is illustrated by applying it to three pedagogically interesting examples: the first is from cosmology, the second from high-energy physics and astronomy, 
both with tractable likelihoods, while the third, with an intractable likelihood, is from epidemiology.\footnote{Code to reproduce all of our results is available on \url{https://github.com/AliAlkadhim/ALFFI}.}
\end{abstract}

\maketitle 

\section{Introduction} 
Simulation-based, or likelihood-free, inference is now ubiquitous in the sciences (see, for example, Refs.\cite{Hermans:2022erv,Brehmer:2021ivt, Cranmer:2019eaq}). Recently, Dalmasso \emph{et al.} introduced likelihood-free frequentist inference,\cite{dalmasso2023likelihoodfree} (\verb|LF2I|) a simulation-based method featuring provable frequentist\cite{Neyman:1937uhy} guarantees. 
Let $\mathcal{D}=\left\{X_i \mid i=1,..., N\right\} $ be the set of ``observable" data sampled from a simulator $F_\theta$ and $D=\left\{x_i \mid i=1,..., N\right\} $ be the set of observed data.
Consider a large, in principle infinite, collection of data-driven statements of the form $\theta \in R(D)$ with parameter  $\theta$ a point in the parameter space of a theoretical model, and $R(D)$ is a data-dependent subset of that space.  Given a function of the data $\lambda(\mathcal{D}; \theta)$, called a \emph{test statistic}, the LF2I approach constructs data-driven statements that are either true or false with the guarantee that over a large collection of such statements a fraction $p \ge \tau$ of them will be true. Parameter subsets that satisfy this condition are called \emph{confidence sets}.


The \emph{confidence level} $\tau$ associated with these subsets is the minimum probability to pick  a true statement, at random,  from the collection. The fraction $p$, which may vary over the parameter space, is called the  \emph{coverage probability}, or coverage for short.
The interesting aspect of \verb|LF2I| is that the guarantee holds for any data sample size.

 Confidence sets and classical hypothesis tests are closely related.
 A classical hypothesis test\cite{Neyman:1937uhy} is a procedure for deciding between two hypotheses: a null hypothesis $H_0$ (parameterized by the parameter $\theta_0$) and an alternative hypothesis $H_1$ (parameterized by parameters that differ from $\theta_0$). For example, we may wish to perform the following test:
\begin{equation}
\label{Eq:simp_hyp}
    H_0: \theta=\theta_0 \quad \text{versus} \quad H_1: \theta \ne \theta_0 .
\end{equation}
 The hypothesis test in Eq. \eref{Eq:simp_hyp} is equivalent to
\begin{equation}
    H_0: \theta \in \Theta_0 \text { versus } H_1: \theta \in \Theta_1 ,
\end{equation}
where $\Theta_0 \cap \Theta_1=\emptyset$, and $\Theta_0$ could be a single parameter point (which defines a \emph{simple hypothesis} or more than one point (thereby defining a \emph{composite hypothesis}). 


A good test statistic is one that provides a clear distinction between the hypotheses we wish to test. When the two hypotheses are simple the likelihood ratio  test statistic is known to be optimal\cite{NeymanPearson},  while experience indicates that the likelihood ratio, or a function thereof, works well even when one or both hypotheses are composite.
Depending on the value of the test statistic, evaluated at a set of observed data $\mathcal{D}=D$, the null hypothesis will either be rejected or will fail to be rejected. If the null hypothesis is rejected, we may choose to accept the alternative.

Assuming that large values of $\lambda=\lambda(\mathcal{D};\theta)$ cast doubt on the null hypothesis $\theta = \theta_0$,  the latter is rejected if the p-value,  $\mathbb{P}(\lambda > \lambda_0 | \theta)$, 
 is less than a given threshold $\alpha$, called the \emph{size} or \emph{significance level} of the test\footnotetext{$\alpha$ is also the threshold at which one is willing to commit a Type I error, the probability of erroneously rejecting the null 
 hypothesis when it is in fact true}.\cite{Algeri:2019lah} For example, if the test statistic is a $\chi^2$ function, large values of $\chi^2$ disfavor the hypothesis $\theta = \theta_0$. If the parameter point $\theta_0$ is not rejected it is added to the confidence set $R(D)$. A confidence set is, therefore, the set of parameter points that have not been rejected at level $\alpha$. 
 The key idea of \verb|LF2I| is to approximate
the p-value with a neural network in such a way that the resulting confidence sets satisfy $p \geq \tau$ or, more realistically, $p \approx \tau$.

The approximated p-value function in the \verb|LF2I| approach is computed for a specific data set $D$, therefore, it cannot be used for other similar data sets with the same sample size. Consequently, it cannot be used to check the coverage explicitly at a given parameter point.  
The \verb|LF2I| approach provides an algorithm to train another neural network to approximate the coverage probability over the parameter space of the theoretical model. Unfortunately, a reliable way to quantify the accuracy of the approximated coverage probability function is not available as is true of the neural network approximation of the p-value. The saving grace, however, is that the quality of the confidence sets can be indirectly assessed by directly computing the coverage at any given point. If the coverage probabilities within the neighborhood of the estimated parameters agree with the desired confidence level $\tau$ to, say, 10\% then one may conclude that the confidence sets, $R(D)$, are satisfactory. 
The motivation for the extension introduced in this paper is the desire to have the p-value do double duty: 1) determine the confidence sets and 2) 
permit
the explicit checking of the coverage using the same neural network.

In practice, we choose to approximate the cumulative distribution function (cdf), $\mathbb{P}(\lambda \leq \lambda_0 | \theta)$, with a neural network. 
Our key idea is to make the latter a function of both the parameter point $\theta$ and the ``observed" test statistic $\lambda_0$. This simple extension makes it possible to apply the approximated cdf to any data set that is similar to the observed data $D$ and of the same sample size. 
Moreover, the cost of approximating the cdf is \emph{amortized} \footnotetext{Meaning that even though the neural network is trained for a specific \emph{kind} of data set $D$, it can be used both for the observed data set and other similar data sets with the same sample size, without the need to re-train the network.} over its subsequent use in constructing confidence sets and in explicitly checking the coverage of these sets. To distinguish the modified \verb|LF2I| from the original, we refer to the former as amortized likelihood-free frequentist inference (\verb|ALFFI|). The \verb|ALFFI| approach is illustrated in three pedagogical examples: the first is from cosmology, the second from high-energy physics and astronomy, and the third from epidemiology. The first two examples feature likelihoods that are tractable, while for the third example the real power of \verb|LF2I| and \verb|ALFFI| is illustrated with a problem in which  the likelihood is intractable.

 The paper is organized as follows. In Sec.\,\ref{sec:ALFFI}, we describe the \verb|ALFFI| approach. This is followed, in Sec.\,\ref{sec:examples}, with the three examples. Section\,\ref{sec:phantom} uses \verb|ALFFI| to infer the parameters of a simple cosmological model that is fitted to Type 1a supernova data, while
 Sec.\,\ref{sec:ON_OFF} applies \verb|ALFFI| to the prototypical signal/background problem in high-energy physics, which in astronomy is known as the On/Off problem.\cite{Li:1983fv}
 For both of these problems, the likelihood is tractable. Section\,\ref{sec:SIR} illustrates the application of \verb|ALFFI| to  a well-known epidemiological model, which, though simple, has an intractable likelihood. The paper ends with a brief discussion in Sec.\,\ref{sec:discussion} and our conclusions in Sec.\,\ref{sec:conclusions}.

\section{Amortized Likelihood-free Frequentist Inference} 
\label{sec:ALFFI}

\subsection{From cumulative distribution function to confidence sets}
A classical hypothesis test is designed to reject hypotheses. Consider the $\alpha$-level hypothesis $\theta = \theta_0$, as in Eq. \eref{Eq:simp_hyp}. This hypothesis is to be rejected if $\mathbb{P}(\lambda \> \lambda_0| \theta_0) < \alpha$. The corollary is that $\theta_0$ is \emph{not}  to be rejected if 
$\mathbb{P}(\lambda > \lambda_0| \theta_0) \geq \alpha$, that is, if
$\mathbb{P}(\lambda \leq \lambda_0| \theta_0) \leq 1 - \alpha \equiv \tau$. The set of points $\theta_0$ that have not been rejected at level $\alpha$, and therefore remain as potentially viable hypotheses for the true value of $\theta$, is by definition a confidence set $R(D)$ at  $100 \tau$\% confidence level (CL). 
The boundary of the confidence set $R(D)$ is determined by the equation 
\begin{align}
    \mathbb{P}(\lambda \le \lambda_0 | \theta_0) & = \tau. 
    \label{eq:CLset}
\end{align} 
By construction, the coverage probability of such sets is

\begin{equation}
\label{Eq:Control_type_1_error}
 \mathbb{P}(\theta \in R({\cal D}) | \theta) \ge \tau.   
\end{equation}
Note that $R(\mathcal{D})$ is a \emph{random} set. Under repeated observations the relative frequency with which these sets include the true value of $\theta$ (ideally) never falls below the stated confidence level $\tau = 1 - \alpha$ whatever the true value of $\theta$. This is a classic example of the \emph{frequentist principle}, whereby in an infinite ensemble of statements, which need not be of  the same kind but which are constructed using the same protocol, there is an \emph{a priori} guarantee that nowhere will the coverage probability fall below the stated confidence level. To the degree that the  probability $\mathbb{P}(\lambda \le \lambda_0 | \theta)$ is accurately modeled, the \verb|LF2I| and \verb|ALFFI| approaches yield random sets that satisfy the frequentist principle.


\subsection{Data Preparation for ALFFI}


The \verb|LF2I| and \verb|ALFFI| approaches are applicable to
any test statistic $\lambda(\mathcal{D}; \theta)$ that is monotonic in the following sense. The test statistic is constructed so that a particular direction in its 1-dimensional space corresponds to hypotheses that are increasingly disfavored. In \verb|ALFFI|, we consider test statistics for which large values correspond to disfavored hypotheses, or equivalently, small values of the 
    $\textrm{p-value} = \mathbb{P}(\lambda > \lambda_0 | \theta=\theta_0)$,
where $\lambda_0 \equiv \lambda(\mathcal{D}=D; \theta_0)$ is the observed value of the test statistic for the specified hypothesis. \verb|ALFFI| (see Sec. \ref{alg:ALFFI}) approximates the probability\cite{Gillespie1983}
\begin{align}
    \mathbb{P}( \lambda \le \lambda_0 | \theta)
    &= \int^{\lambda_0} d Y \int d{\cal D} \, \delta(Y - \lambda({\cal D}, \theta)) \, p({\cal D} | \theta) ,
        \label{eq:cdf}
\end{align}
for any data set $\mathcal{D}$
sampled from a statistical model $p({\cal D} | \theta)$, which may or may not be tractable. However, as in the \verb|LF2I| 
approach,\cite{dalmasso2023likelihoodfree} it is assumed that one has access to a large collection $\mathbb{S}$ of simulated pairs $({\cal D}, \theta) \in \mathbb{S}$ where for every point $\theta$ sampled from any convenient prior $\pi_\theta$ a \emph{single} instance of a data set ${\cal D}$ is simulated with the same characteristics, including sample size, as the real data $D$.  
In contrast to \verb|LF2I|, in \verb|ALFFI| a second collection of data sets $\mathbb{S}^\prime \ni {\cal D}^\prime$ is created from $\mathbb{S}$ by randomly shuffling its data sets  so that they are de-correlated with respect to the parameter points in $\mathbb{S}$. 
The data sets in $\mathbb{S}^\prime$ serve as instances of ``observed" data sets. 

For every parameter $\theta$ in $\mathbb{S}$, we compute two values of the test statistic, namely, $\lambda({\cal D}, \theta)$ and $\lambda_0 = \lambda({\cal D}^\prime, \theta)$, as well as the discrete variable $Z$ which is unity if $\lambda({\cal D}, \theta)  \leq \lambda({\cal D}^\prime, \theta)$ and zero otherwise. This procedure results in a large collection of triplets $\mathcal{T} = \{ (Z, \lambda_0, \theta )\}$, which constitute the training data. In \verb|LF2I|, the observed test statistic---the second component of the triplet---is computed using a \emph{fixed} data set ${\cal D}^\prime = D$, namely, the one actually observed, while \verb|ALFFI| uses the data sets ${\cal D}^\prime \in \mathbb{S}^\prime$. 

\subsection{Approximating the cumulative distribution function}
\label{sec:cdf}

The cumulative distribution function $\mathbb{P}( \lambda \le \lambda_0 | \theta)$ is the expectation value $\mathbb{E}(Z | \lambda_0, \theta)$ of the discrete variable $Z$, a fact that suggests a straightforward way to approximate the cdf for a \emph{fixed} data set $D$: histogram the parameter points $\theta$, thereby yielding the histogram $\mathbb{H}_1$, and using the same bins as $\mathbb{H}_1$ histogram the parameter points again, but this time weighted by $Z$, yielding the histogram $\mathbb{H}_Z$. The ratio $\mathbb{H}_Z / \mathbb{H}_1$ approximates $\mathbb{E}(Z |\lambda_0, \theta)$. 


Following \verb|LF2I|, a more convenient, and one hopes better, approximation of $\mathbb{E}(Z |\lambda_0, \theta)$ is created with a deep neural network (DNN) trained (that is, fitted) by minimizing the empirical risk
\begin{align}
    E(\boldsymbol{\omega}) & = \frac{1}{N} \sum_{i=1}^N \mathcal{L}(t_i, f_i),
    \label{eq:erisk}
\end{align}
where $\mathcal{L}(t, f)$ is a loss function, $f(\bold{x}; \boldsymbol{\omega})$ is a DNN with inputs $\bold{x}$ and free parameters $\bold{\omega}$, and $t$ denotes known \emph{targets}. The empirical risk or average loss, Eq. \eref{eq:erisk}, is a Monte Carlo approximation of the risk functional
\begin{align}
    E[f] & = \int \int  \mathcal{L}(t, f) \, p(\bold{x}, t) \, d\bold{x} \, dt,
    \label{eq:risk}
\end{align}
where $p(\bold{x}, t)  = p(t | \bold{x}) \, p(\bold{x})$ is the (typically unknown) probability distribution of the data $\bold{x}, t$. From the calculus of variations, the function $f$ that minimizes Eq. \eref{eq:risk} is the solution of
\begin{align}
    \int \frac{\partial \mathcal{L}}{\partial f} \, p(t | \bold{x}) \, dt & = 0, 
    \label{eq:solution}
\end{align}
assuming that $p(\bold{x}) > 0$ $\forall \, \bold{x}$. 
In the examples below, and following \verb|LF2I|, we use the quadratic loss $\mathcal{L}(t, f) = (t - f)^2$, which, using Eq. \eref{eq:solution}, leads to the well-known result\cite{bayesnet1, bayesprob2} 
\begin{align}
    f(\bold{x}; \boldsymbol{\omega}^*) & = \int t \, p(t | \bold{x}) \, dt \equiv \mathbb{E}_t[t \mid \bold{x}],
    \label{eq:fx}
\end{align}
where $\boldsymbol{\omega}^*$ are the best-fit parameters of the neural network model. 
Setting $\bold{x} = \lambda_0, \theta$ and the targets $t = Z$ in Eq. \eref{eq:fx} yields 
\begin{align}
\label{Eq:NN_appeox_EZ}
    f(\lambda_0, \theta; \boldsymbol{\omega}^*) & \approx \mathbb{E}\left[ Z \mid \lambda_0, \theta \right],\nonumber\\
    &= \mathbb{P}( \lambda \le \lambda_0 | \theta),
\end{align}
that is, it yields the quantity that we wish to approximate.

One of the key virtues of the \verb|LF2I| and \verb|ALFFI| approaches, as is evident in the result in Eq. \eref{Eq:NN_appeox_EZ}, is that the neural network $f$ is conditioned on $\theta$, which means that it is independent of the prior $\pi_\theta$. The form of the prior affects only the accuracy of the approximation. The accuracy of the approximation will be greatest where the density of the prior is greatest.

\section{Results}
\label{sec:examples}
The \verb|ALFFI| approach is illustrated in the following three diverse examples: the first is from cosmology, the second from high-energy physics and astronomy, and the third from epidemiology. Our choice of the particular problem in each field highlights  typical statistical inference problems in each field. We demonstrate that the \verb|ALFFI| method yields valid multi-parameter confidence sets for all three examples in a computationally efficient manner, both in cases where the likelihood is tractable (the first two examples), and where the likelihood is intractable (the third example). We also demonstrate the use of different test statistics, demonstrating the compatibility with binned and un-binned analyses. Table \ref{table:examples} summarizes key attributes of each example.

\begin{table}[h]
\caption{\emph{Cosmological Model}: $x_i \pm \sigma_i$ and $z_i$ are the measured distance moduli and redshifts, respectively, while ${\cal H}_0$ and $n$ are the model parameters. \emph{Signal/Background}: $N$ and $M$ are the observed counts for signal and background, respectively, while $n$ and $m$ are the expected signal and background counts, respectively. $\mu$ and $\nu$ are the unknown signal mean (parameter of interest) and unknown background mean (nuisance parameter), respectively. \emph{SIR Model}:  $\{ x_i \}$ are the 13 observed counts of infected children on the 13 days of observation, and $\alpha$ and $\beta$ are the model parameters. CTMC is a continuous time Markov chain model of the epidemic.}
\centering
\noindent
\small
\begin{tabular}{|p{\linewidth/7}|p{\linewidth/11}|p{\linewidth/16}|p{\linewidth/4}|p{\linewidth/4}|p{\linewidth/10}|p{\linewidth/14}|}
\hline 
  Example & {Observed \newline  data ($D$)} &  $\theta$ & Priors $\pi_\theta$  & {$\mathcal{D} \sim F_\theta$}  & Tractable $\mathcal{L}$ ? & $\lambda$ \\
\hline
  {Cosmological\newline Model} & $x_i \pm \sigma_i, \newline z_i$ & $\mathcal{H}_0,\newline n$  & {$\frac{{\cal H}_0}{100} \sim \text{Unif}(0.66, 0.76)$, \newline $n \sim \text{Unif}(0.05, 0.65)$} & $x_i \sim \mathcal{N}(\mu, \sigma_i)$ & Yes &  Eq. \eref{Eq:lambda_cosmo} \\
\hline
  {Signal/ \newline background} & {$N=3$, \newline $M=7$} & $\mu$,\newline $\nu$ & {$\mu \sim \text{Unif}(0,20)$,  \newline$\nu \sim \text{Unif}(0,20)$} & {$n \sim \text{Poisson}(\theta+\nu)$,  \newline $m \sim \text{Poisson}(\nu)$} & Yes & Eq. \eref{Eq:lambda_on_off}  \\
\hline
  SIR\newline Model & $x_i$ &  $\alpha,\newline \beta$ & {$\alpha\sim \text{Unif}(0.1,0.9)$,  \newline $\frac{\beta \times10^{3}}{5} \sim \text{Unif}(0.25,0.65)$} & $\{ x_i\} \sim \text{CTMC}(\theta)$ & No & Eq. \eref{Eq:lambda_SIR}   \\
\hline
\end{tabular}
\label{table:examples}
\end{table}

\subsection{Example 1: Cosmological Model}
\label{sec:phantom}

 In the late 1990s, fits of cosmological models to Type 1a supernova data led to the conclusion that the expansion of the universe is accelerating.\cite{Riess1998,Perlmutter1999} The fits then, as now, were 
 performed using tractable likelihoods, typically a multivariate normal. 
 In this example, we fit a cosmological model to the 
 Union 2.1 data compilation of the Supernova Cosmology Project\cite{Suzuki2012} via maximum likelihood and also with \verb|ALFFI|. The Union 2.1 data set comprises  measured distance moduli, $x \pm \sigma$, and redshifts, $z$, for 580 Type 1a supernovae.
 Given the size of the data sample, it is expected that accurate confidence sets for the cosmological parameters can be constructed using standard methods such as maximum likelihood. \verb|ALFFI| is therefore not needed for this problem; it is simply used to showcase the algorithm.

Our cosmological model is defined by the 
equation of state 
\begin{align}
    {\cal P} & = -b a^n \Omega, 
    \label{eq:eos}
\end{align}
where $n$ and $b$ are free parameters, and
$a(t)$, $\Omega(a)$, and ${\cal P}$ are the dimensionless universal scale factor, the dimensionless energy density, and the dimensionless pressure, respectively, and $t$ is the time since the Big Bang. 
For $n > 1$ and $a \ll 1$, the equation of state is that of a pressureless dust of particles as in the $\Lambda$CDM model.\cite{Peebles2003} However, at later times the energy density becomes dominated by so-called phantom energy.\cite{PhantomPhysRevLett.91.071301} Our model
is consistent with the
Friedmann–Lemaître–Robertson–Walker
(FLRW) metric with zero curvature,  and the Friedmann equations,\cite{Peebles2003}
\begin{align}
\left(\frac{1}{a}\frac{da}{dt}\right)^2 & = {\cal H}_0^2  \Omega(a) \label{eq:f1}\\
     \textrm{and } \, a\frac{d\Omega}{da} & = -3 (\Omega + {\cal P}), \label{eq:f2}
\end{align}
are assumed to hold, where ${\cal H}_0$ is the Hubble constant. The first Friedmann equation, Eq. \eref{eq:f1}, incorporates the convention $a(t_0) = 1$ at the present epoch $t_0$. By definition, the Hubble constant is the present value of the Hubble parameter ${\cal H}(a) = a^{-1} da/dt$, which implies $\Omega(1) = 1$ for all cosmological models. 

 Combining Eqs.\eref{eq:eos} and \eref{eq:f2}, integrating the latter, and imposing the constraint $\Omega(1) = 1$, yields
\begin{align}
    \Omega(a) & = \exp[3 b(a^n - 1)/n] \, / \, a^3,
\end{align}
for the dimensionless energy density.
This model is defined by the three parameters $n$, $b$, and ${\cal H}_0$, but we reduce it to a 2-parameter model by choosing $b = n / 3$.
From the first Friedmann equation, Eq. \eref{eq:f1}, the energy density $\Omega(a)$ and the dimensionless time ${\cal H}_0 t$ are related as follows
\begin{align}
    {\cal H}_0 t & = \int_0^a \frac{dy}{y \sqrt{\Omega(y)}} , \nonumber\\ 
          & = \sqrt{e} \, 2^{\frac{3}{2 n}}  \gamma\left(\frac{3}{2 n}, \frac{a^n}{2}\right) \, / \, n , 
\end{align}
where $\gamma(a, x) = \int_0^x \, t^{a - 1} \, e^{-t} \, dt$ is the 
lower incomplete gamma function. 
Setting $a = 1$ yields
the dimensionless age of the universe
\begin{equation}
    {\cal H}_0 t_0 = \sqrt{e} \, 2^{\frac{3}{2 n}}  \gamma\left(\frac{3}{2 n}, \frac{1}{2}\right) \, / \, n .
\end{equation}

\begin{figure}[h!]
\centering
\includegraphics[width=0.8\textwidth]{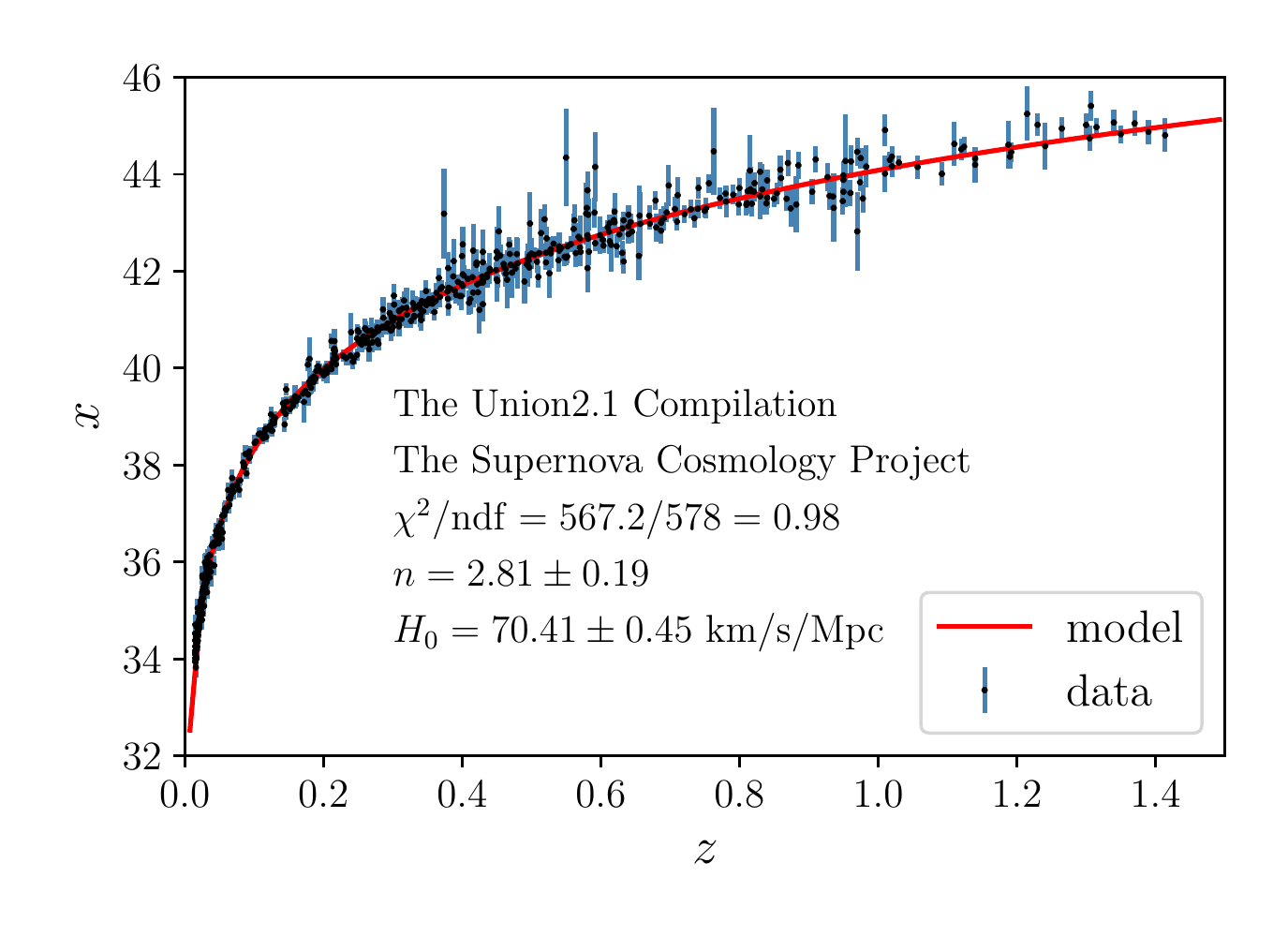}
\caption{Example 1 (cosmology): Fit of 2-parameter cosmological model to Union 2.1 Type 1a data set. Note the excellent fit, which with a $\chi^2$ per number of degrees of freedom (ndf) of 0.98  is on par with that of the $\Lambda$CDM model.}
\label{fig:phantom_fit}
\end{figure}

If the small correlations between the 580 Typa 1a data points are neglected, the likelihood for these data is a diagonal multivariate normal. Maximizing this likelihood with respect to its parameters is equivalent to minimizing the function 
\begin{equation}
    \chi^2  =  \sum_{i=1}^N \left( \frac{x_i - \mu(z_i, \theta)}{\sigma_i}\right)^2 ,
    \label{eq:chi2}
\end{equation}
where $\mu(z, \theta)$, the \emph{distance modulus}\cite{Peebles2003}---an astronomical measure of distance, is given by
\begin{equation}
    \mu(z, \theta) = 5 \log_{10}[(1 + z) \, \sin(\sqrt{-\Omega_K} \, u(z, \theta))\, / \, \sqrt{-\Omega_K}] + 5 \log_{10}(c \, / \, {\cal H}_0 / 10^{-5} \textrm{Mpc}).
\end{equation}
The quantity
$c$ is the speed of light in vacuum in km/s, $\Omega_K$ is the curvature parameter,
 and 
\begin{equation}
    u(z, \theta) = \int_{1/(1+z)}^{1} \frac{da}{a^2\sqrt{\Omega(a)}},\nonumber\\
 = 2^{\frac{1}{2 n}} \left[\gamma\left(\frac{1}{2 n}, \frac{1}{2}\right) - \gamma\left(\frac{1}{2 n}, \frac{\left(1 + z\right)^{-n}}{2}\right)\right] \sqrt{e} / \, n ,
 \label{eq:u}
\end{equation}
is a dimensionless function.
With $\Omega_K \rightarrow 0$, the distance modulus simplifies to 
\begin{align}
    \mu(z, \theta) 
    & = 5 \log_{10}\left[ (1 + z) c \, u(z, \theta) \, /\,  {\cal H}_0 \right] + 25 . 
    \label{eq:mu}
\end{align}
When the model is fitted to the Union 2.1 data set by minimizing Eq. \eref{eq:chi2} an 
excellent fit is obtained, as shown in Fig.\,\ref{fig:phantom_fit}.
For $t < t_0$, the time dependence of the scale factor $a(t)$ is similar to that of the standard $\Lambda$CDM model. But the model exhibits 
a future singularity (a Big Rip) characterized by the condition $a \rightarrow \infty$ at a \emph{finite} time given by
\begin{equation}
    {\cal H}_0 t_\textrm{rip} = \sqrt{e} \, 2^{\frac{3}{2n}} \, \Gamma\left(\frac{3}{2n}\right) \, / \, n,
\end{equation}
that is, at 
\begin{align}
    t_\textrm{rip} & = \frac{\Gamma\left(\frac{3}{2n}\right)}{\gamma\left(\frac{3}{2 n}, \frac{1}{2}\right) } \, t_0 .
\end{align}

We now apply \verb|ALFFI| to the same problem using the test statistic 
\begin{equation}
\label{Eq:lambda_cosmo}
    \lambda(D, \theta)  = \sqrt{\frac{\chi^2}{N} 
    },
\end{equation}
where $\chi^2$ is defined by the Eqs.\,(\ref{eq:chi2}), (\ref{eq:u}), and (\ref{eq:mu}). The form of the test statistic 
is chosen
so that it is  ${\cal O}(1)$. 
The boundary of the associated $100\tau$\% confidence set is given by Eq. \eref{eq:CLset}, which requires a good approximation to
the cumulative distribution function
$\mathbb{P}(\lambda \le \lambda_0 | \theta)$. The latter is approximated in two ways: with histograms and with a deep neural network (DNN) as described in 
Sec.\,\ref{sec:cdf}. The 2D histograms have 10 bins in  both dimensions $n$ and ${\cal H}_0$ with the parameter $n$ scaled down by a factor 10 and ${\cal H}_0$ scaled down by a factor 100. Since the same simulated Type 1a data are used for both the histogram and the DNN-based approximations, the parameters are scaled so that all inputs to the DNN are ${\cal O}(1)$, which is considered good practice. 

The DNN is 1,781-parameter fully-connected feed-forward neural network, 
\begin{align}
    f(\boldsymbol{x}) &= \text{sigmoid}(b_5 + \boldsymbol{w}_5 h( \boldsymbol{b}_4 + \boldsymbol{w}_4 h( \boldsymbol{b}_3 + \boldsymbol{w}_3 h( \boldsymbol{b}_2 + \boldsymbol{w}_2 h( \boldsymbol{b}_1 + \boldsymbol{w}_1 h(  \boldsymbol{b}_0 + \boldsymbol{w}_0 \boldsymbol{x}  )    )    )   )   )) ,
\end{align}
with 3 input features, $\boldsymbol{x} = \{ \lambda_0, n, {\cal H}_0 \}$,  5 hidden layers with 20 nodes each, and a single output. The elements of the matrices $\boldsymbol{b}_i$ and $\boldsymbol{w}_i$ are the free parameters of the DNN, and
    the non-linear function, $h$, at each hidden node is a ReLU applied element-wise, that is, to every element of its matrix input. The output node is a sigmoid that constrains the output to lie within the unit interval as befits a probability. The DNN is trained as described in Sec.\,\ref{sec:ALFFI} using PyTorch\cite{pytorch}.

In practice, at each gradient descent step in the ``landscape" defined by the empirical risk (that is, average loss), Eq. \eref{eq:erisk}, the gradient is computed from a batch of data of size $K = 50 \ll N$ randomly sampled from $N = 250,000$ sets of 580 simulated distance moduli. The steps are computed using the Adam\cite{kingma2017adam} optimizer with a fixed learning rate of $10^{-3}$. As the training proceeds, the network with the smallest average loss is saved. The average loss is computed using a validation data set of size 5,000 that is not used by the optimizer. The training stops if after 50,000 steps no network is found with a smaller validation loss than the saved network. The best network as defined by this training protocol is found after about 100,000 iterations. If one defines an epoch to be $N / K $ iterations, which for this example is 5,000, then $100,000$ iterations corresponds to 20 epochs of training.
The simulated sets of 580 distance moduli are sampled from 580 independent normal distributions using the standard deviations taken from the observed data.

\begin{figure}[h!]
\centering
\includegraphics[width=0.6\textwidth]{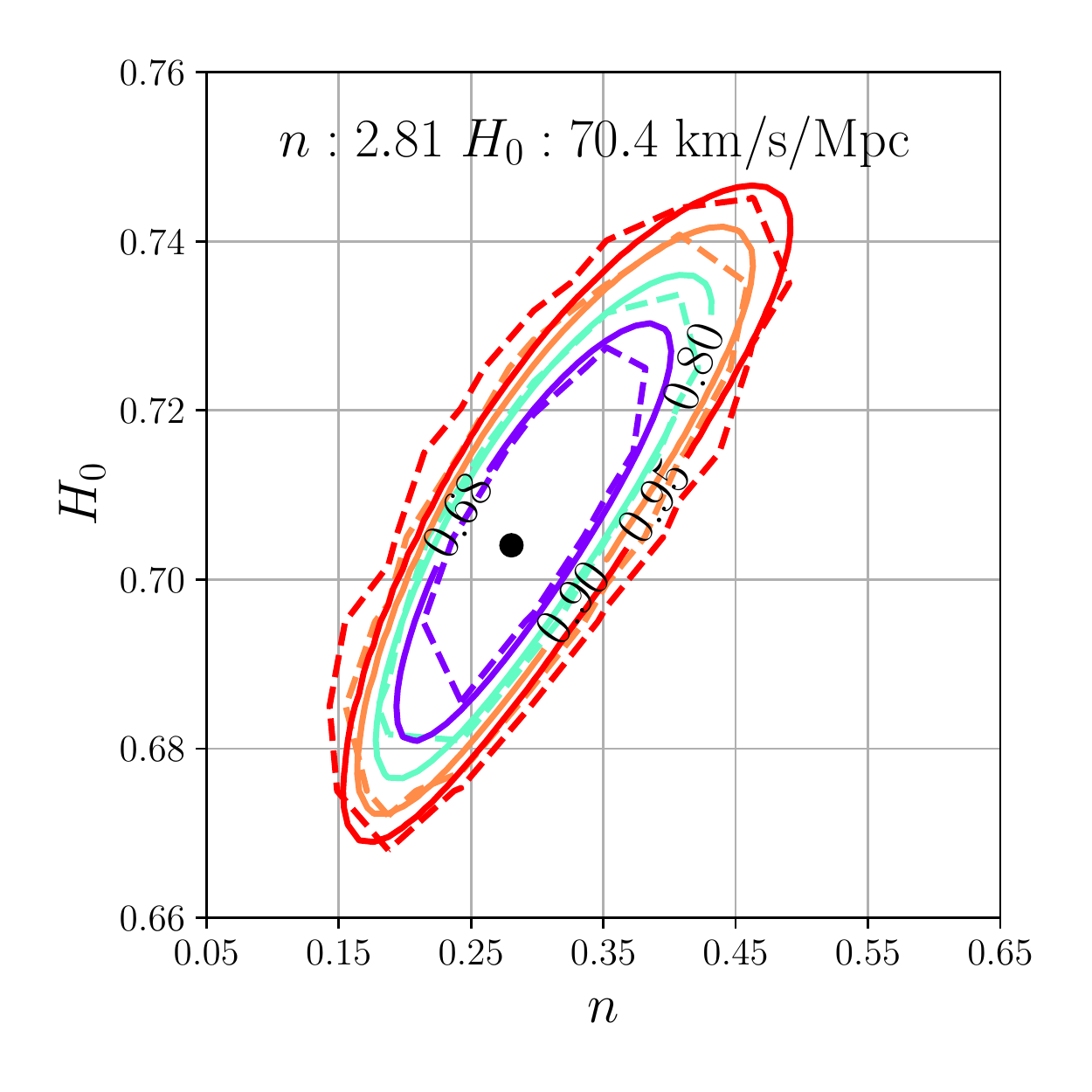}
\caption{Example 1 (cosmology): Confidence sets $R(D)$ for $\tau = 0.68, 0.80, 0.90$, and $0.95$. (\emph{dashed lines}) Boundaries of confidence sets, $R(D)$, defined by $\mathbb{P}(\lambda \le \lambda_0 | \theta) = \tau$ computed using the histogram-based approximation of the cdf.
(\emph{solid lines}) Boundaries of confidence sets computed using the DNN-based approximation of the cdf. (\emph{black dot}) Location of the minimum of $\mathbb{P}(\lambda \le \lambda_0 | \theta)$, computed with the DNN approximation, which is taken to be the best-fit point.}
\label{fig:phantom_results}
\end{figure}

Approximating 
$\mathbb{P}(\lambda \le \lambda_0 | \theta)$ using histograms and using \verb|ALFFI| leads to the confidence sets shown in Fig.\,\ref{fig:phantom_results}.
The best-fit values of the cosmological parameters $\theta = \{ n, {\cal H}_0 \}$ are taken to be the location of the minimum of $\mathbb{P}(\lambda \le \lambda_0 | \theta)$, which as indicated in Fig.\,\ref{fig:phantom_results} agrees with the values obtained from the likelihood fit.

\begin{figure}[h!]
\centering
\includegraphics[width=\textwidth]{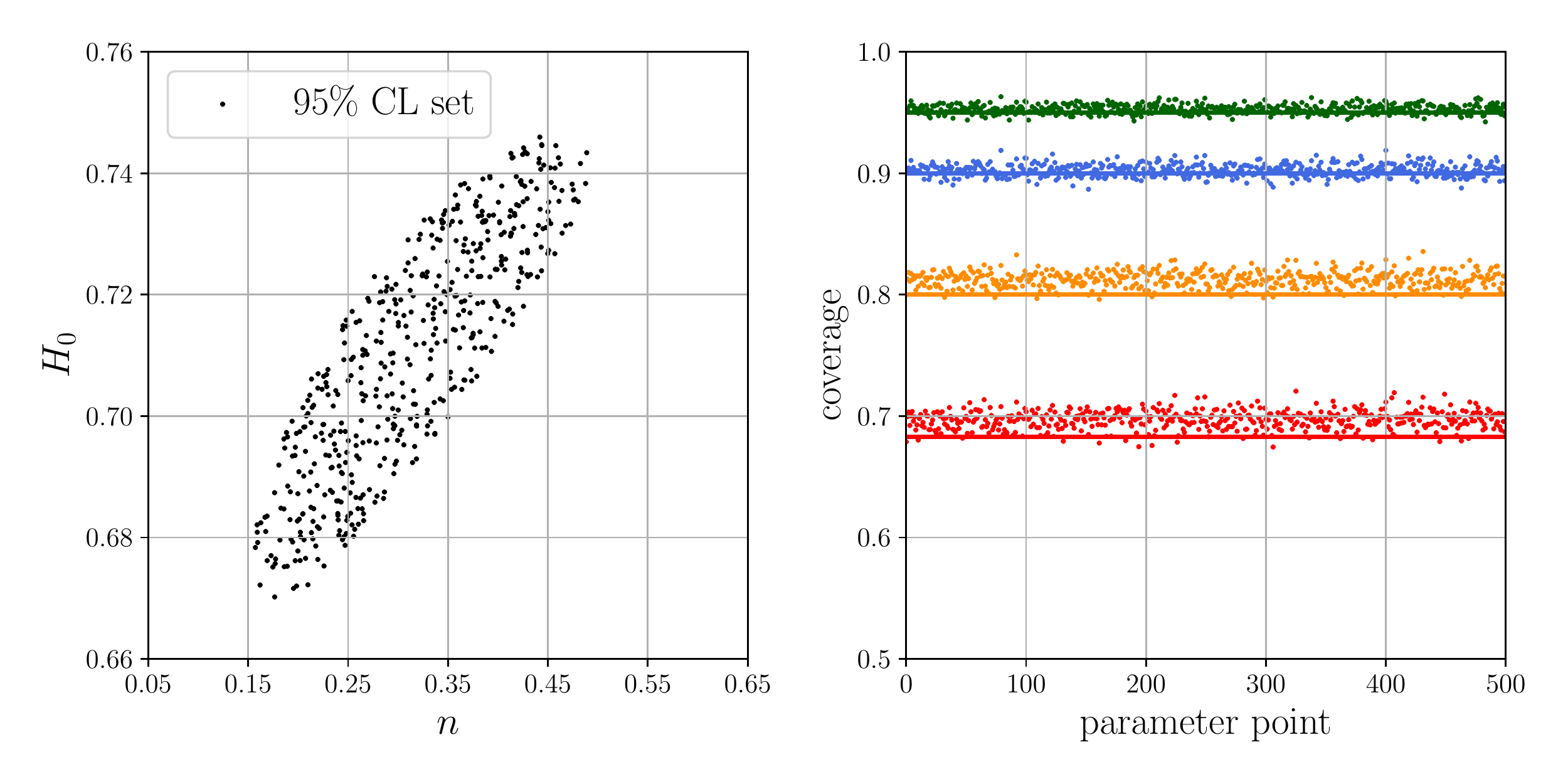}
\caption{Example 1 (cosmology):  (\emph{left}) 500 points, from the 95\% CL confidence set for the Type 1a data, at which the coverage has been computed. (\emph{right}) the dots are the coverage probabilities for each of the 500 points in the cosmological model $n, {\cal H}_0$ parameter space and fhe horizontal lines are the values of the confidence levels $\tau$.}
\label{fig:phantom_cov}
\end{figure}

In the \verb|LF2I| approach, the  coverage  over the parameter space is checked by modeling the coverage probability with another neural network as a function of the parameters of the theoretical model. There are pros and cons to that approach. It is certainly convenient to have a functional approximation of the coverage probability as a function of the parameter space point because one can then estimate the coverage at any given point, not only at points for which there are sufficient simulated data. Unfortunately, however, as is true of most machine learning models, a reliable estimate of the accuracy of the trained machine learning model is not available.  In \verb|ALFFI|, the coverage is checked explicitly by direct enumeration at all points for which there are sufficient data. Since the problem consists of counting how often a particular statement is true, the problem is binomial; therefore, a reliable estimate of the accuracy of the coverage calculation is easy to compute. On the other hand, the coverage is available only at the parameter points for which there are sufficient simulated data. In this example, for a given parameter point, $T = 4,000$ sets of 580 Type1a supernovae data are simulated. For each data set, a test statistic, $\lambda_0$, is computed. If $\mathbb{P}(\lambda \le \lambda_0 | \theta) \le \tau$ then, by definition, $\theta$ lies within the confidence set associated with $\lambda_0$. If $S$  is the number of times this statement is true over the collection of $T$ simulated data sets, then the coverage probability is $p  \pm \sqrt{p (1 - p) / T}$, where $p = S / T$. If the confidence sets produced by \verb|ALFFI| are accurate then we should find that $p \approx \tau$. This calculation is performed at 500 randomly selected points within the 95\% CL set associated with the observed Type 1a data, as shown in the left panel of Fig.\,\ref{fig:phantom_cov}.
The right panel shows that the coverage of the confidence sets computed using \verb|ALFFI| do indeed have correct coverage and are, therefore, in that sense accurate.

\subsection{Example 2: Signal/Background or ON/OFF Model}
\label{sec:ON_OFF}

The cosmological example served to illustrate the \verb|ALFFI| algorithm, but, as noted, \verb|ALFFI| is not really needed  because the cosmological data are numerous, the likelihood function is tractable, and parameter estimation via maximum likelihood works well. Our second example addresses 
the signal/background problem in high-energy physics (see, for example, 
Ref.\cite{Lyons:2008zz}),
which in astronomy
is referred to as the On/Off problem.\cite{Li:1983fv} We choose an example in which the likelihood is tractable but the data are sparse and, consequently, asymptotic methods may not be reliable.\cite{Algeri:2019lah} We demonstrate that the ALFFI method, which is based on the Neyman construction,  yields valid results even when the regularity conditions that underpin Wilks' theorem\cite{WilksThm} and its variants \cite{Cowan2011} are violated.

The signal/background problem in high-energy physics and astronomy is as follows. An observation is made, for given period of time, which consists of counting $N$ events: typically, photons in astronomy and particle collisions in high-energy physics. The count is potentially a sum of counts from signal and background sources. A second independent observation is made for the same duration (in the simplest case) where by design the background has the same characteristics as in the first observation but no signal is present. The second observation yields a count $M$. It is generally assumed that the likelihood function for the data $D =\{ N, M \}$ is the product of two Poisson distributions, 
\begin{align}
    \mathcal{L} (D ; \mu, \nu) & = \frac{(\mu + \nu)^N \exp(-(\mu + \nu))}{N!} \,\frac{\nu^M\exp(-\nu)}{M!},
\label{Eq:OnOffModel}
\end{align}
where $\mu$ and $\nu$ are the mean signal and background counts, respectively. In the signal/background problem the parameter of interest is $\mu$, while $\nu$ is a \emph{nuisance parameter}. We shall comment on how one might deal with such parameters in the discussion.

Our specific example is from the first experiment to search for neutron-antineutron oscillations using free neutrons,\cite{CRISP:1985fte} which took place in the 1980s at the Institut Laue-Langevin (ILL) in Grenoble, France. Neutron-antineutron ($n\bar{n}$) oscillations are predicted by many proposed theories of physics beyond the Standard Model of particle physics.\cite{Phillips:2014fgb} For our purposes it suffices to note that if $n\bar{n}$ oscillations can occur then a pure neutron state, when observed at time $t \ll$ than the mean neutron lifetime, will 
be observed with probability
\begin{align}
    P_t & = \left(\frac{\epsilon^2}{\epsilon^2 + \Delta E^2} \right)\sin^2((\epsilon^2 + \Delta E^2)^{1/2} t),\nonumber\\
    & \approx \left(\frac{\epsilon^2}{\Delta E^2} \right)\sin^2( \Delta E \, t), 
    \label{eq:Pt}
\end{align}
 as an antineutron state, where $2 \Delta E$ is the difference in neutron and antineutron energies in external fields and $\epsilon = \tau_{n\bar{n}}^{-1}$ is the energy characteristic of whatever new physics is responsible for the oscillations; $\tau_{n\bar{n}}$ is referred to as the oscillation time. We use units in which $\hbar = 1$. 
The experimental conditions are such that $\Delta E \gg \epsilon$, which justifies the 
approximation in Eq. \eref{eq:Pt}. Furthermore, in the Grenoble experiment the quasi-free condition $\Delta E t \ll 1$ could be realized, leading to a transition probability, 
\begin{align}
    P_t & \approx (t \, / \, \tau_{n\bar{n}})^2,
\end{align}
independent of the energy perturbation $\Delta E$ arising from the neutron and antineutron interactions with the ambient magnetic field. In the quasi-free condition $N = 3$ events were recorded in this experiment.

The background in the Grenoble experiment was directly measured by applying a magnetic field  to suppress the transition probability $P_t$ by making $\Delta E$ large enough. This condition yielded $M = 7$ events. The maximum likelihood estimate of the signal is $\hat{\mu} = N - M = -4$ events. However, since $\mu \geq 0$, we choose to take the best estimate of the signal in such experiments to be 
\begin{equation}
     \hat{\mu} =\left\{
    \begin{array}{ll}
        N-M & \text{ if } \quad  N>M \\
        0 & \quad \textrm{ otherwise.}
    \end{array}
    \right. 
    \label{eq:muhat}
\end{equation}
The sparsity of the data in the Grenoble experiment and our choice of signal estimate explicitly violate two of the regularity conditions for standard asymptotic results to hold\,\cite{Algeri:2019lah}: 
the data should be sufficiently numerous and estimates must not lie on the boundary of the parameter space.
The violation of these conditions, however, is not a problem for \verb|LF2I| and \verb|ALFFI|. 

To construct confidence sets in the parameter space of $\theta = \{ \mu, \nu \}$, we use the test statistic
\begin{equation}
    \lambda(D, \theta)  = -2 \log \left[  \frac{\mathcal{L}(D ; \mu, \nu)}{\mathcal{L}(D ; \hat{\mu}, \hat{\nu})} \right],
    \label{Eq:lambda_on_off}
\end{equation}
where $\hat{\mu}$ is given by Eq. \eref{eq:muhat} and $\hat{\nu}(\mu)$ by
\begin{equation}
     \hat{\nu} =\left\{
    \begin{array}{ll}
        M & \text{ if } \quad  \hat{\mu} = N - M \\
        (M+N)/2 & \quad \textrm{ otherwise.}
    \end{array}
    \right. 
    \label{eq:nuhat}
\end{equation}
The cdf, $\mathbb{P}(\lambda \leq \lambda_0 | \theta)$, was again approximated with a fully-connected feed-forward DNN with 3 input features $\mathbf{x} = \{ \lambda_0, \mu, \nu \}$, 6 hidden layers with 12 nodes each, and a single output, estimating $\mathbb{E} \left[ Z\mid\lambda_0,\mu,\nu \right]$. The activation function at each hidden node is a PReLU, \cite{PReLU} and the network was trained with the Adam optimization algorithm with a fixed learning rate of $6 \times 10^{-4}$. The training set is composed of $10^7$ examples, which were used in batches of size $5\times 10^3$, for the duration of $10^{5}$ iterations, that is, for 50 epochs. A batch normalization \cite{batchnorm} layer was added after every hidden layer as a regularization technique.

\begin{figure}[h!]
    \centering
\includegraphics[width=0.6\textwidth]{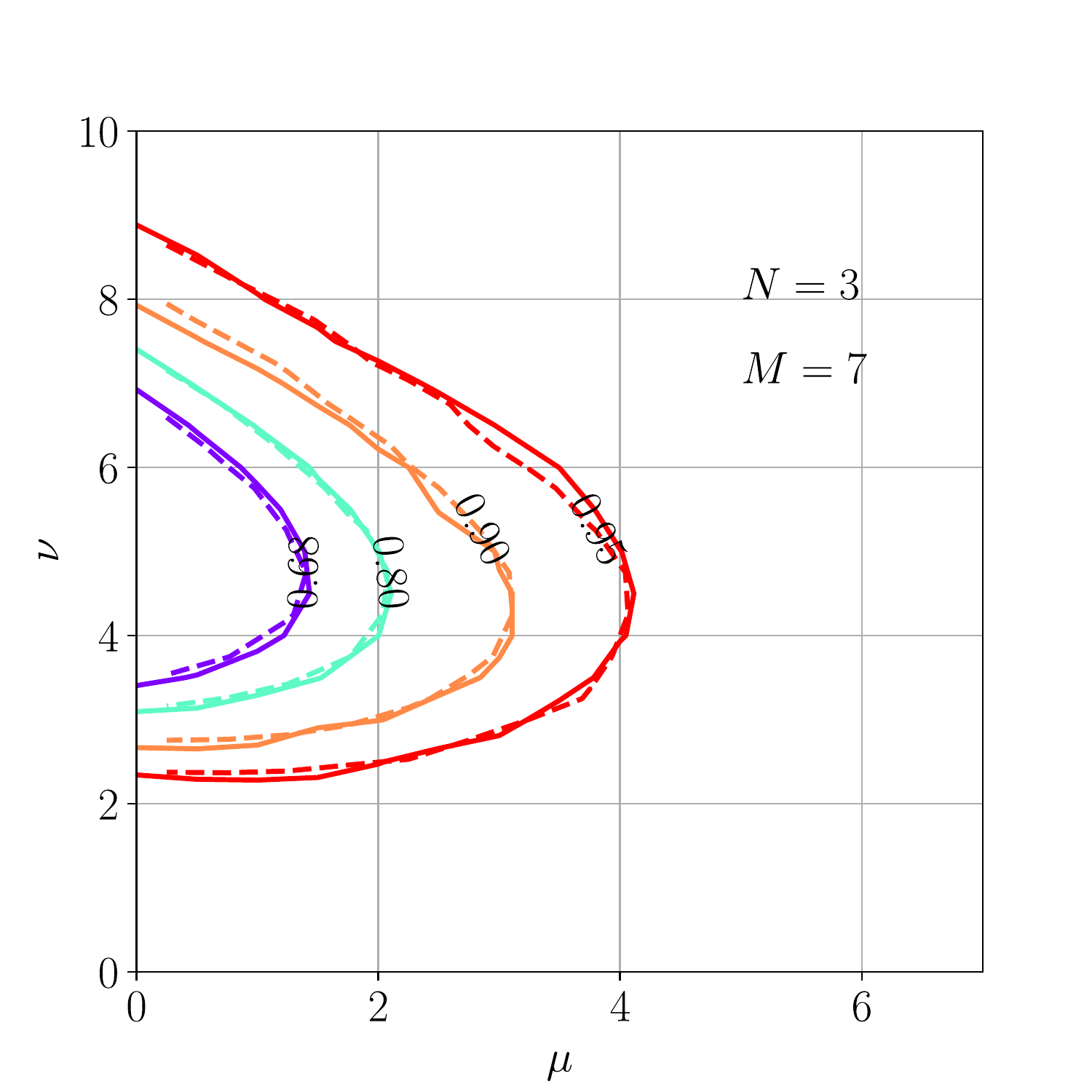}
\caption{Example 2 (signal/background): Confidence sets $R(D)$ for $\tau = 0.68, 0.80, 0.90$, and $0.95$. (\emph{dashed lines}) Boundaries of confidence sets, $R(D)$, defined by $\mathbb{P}(\lambda \le \lambda_0 | \theta) = \tau$ with the histogram-based approximation of the cdf.
(\emph{solid lines}) Boundaries of confidence sets computed using the DNN-based approximation of the cdf. }
\label{fig:LAMBDA_D_AS_DATA}
\end{figure}

\begin{figure}[h!]
\centering
\includegraphics[width=\textwidth]{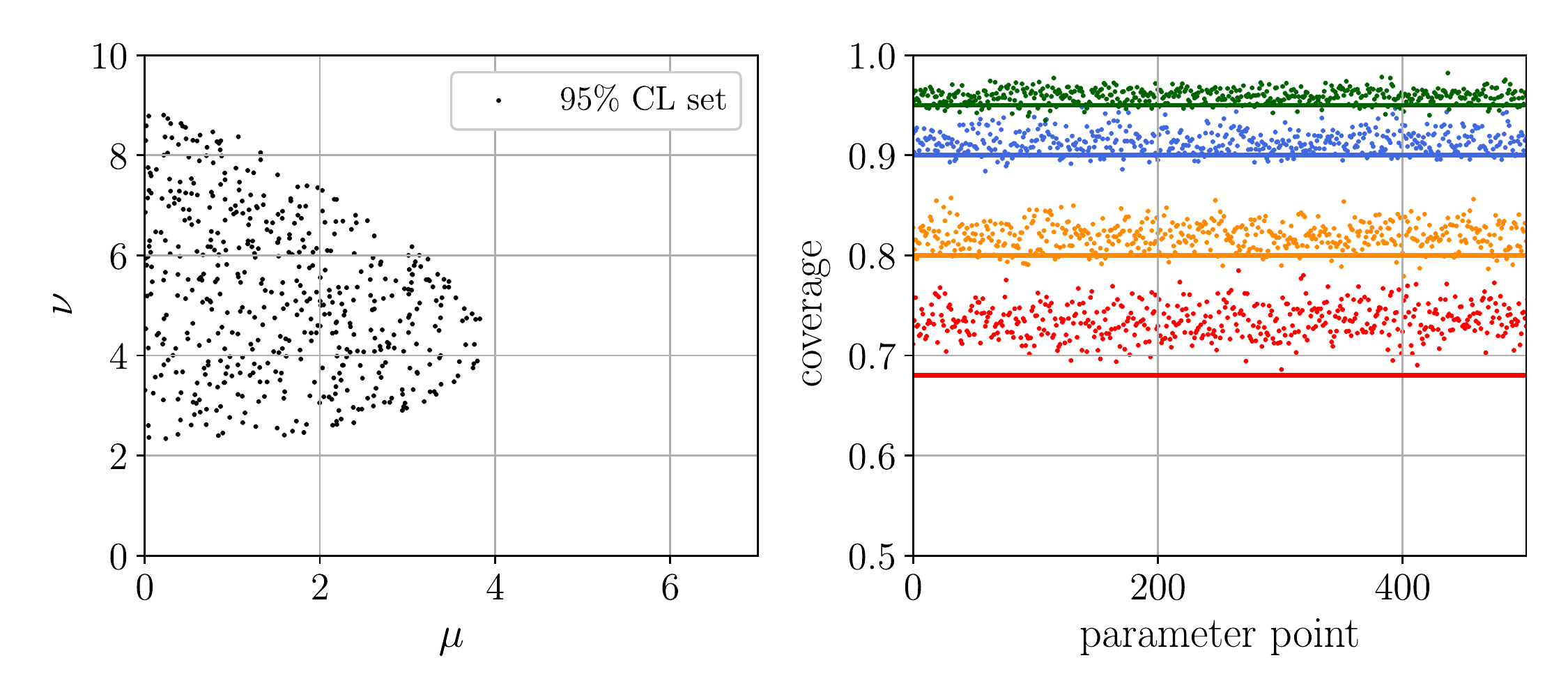}
\caption{Example 2 (Signal/Background): (\emph{left}) 500 points, from 95\% CL confidence set for the Grenoble data,\cite{CRISP:1985fte} at which the coverage has been computed. 
(\emph{right}) the coverage probabilities for each of the 500 points in the signal/background $\{ \mu, \nu \}$ parameter space. 
}

\label{fig:ONOFFcov}
\end{figure}

The DNN was used to compute the confidence sets shown in Fig.\,\ref{fig:LAMBDA_D_AS_DATA} and the associated coverage probabilities shown in Fig.\,\ref{fig:ONOFFcov}. The fact that the coverage probabilities are within about 10\% of the confidence levels confirms the accuracy of the confidence sets obtained with \verb|ALFFI|.

\subsection{Example 3: SIR Model}
\label{sec:SIR}

In the cosmology and signal/background examples, the likelihood functions are tractable. In our third example,  from the field of epidemiology,  the likelihood is intractable. \cite{andersson2012stochastic} Therefore, this example is one for which
\verb|LF2I| and \verb|ALFFI| are the most useful.

\begin{figure}[h!]
\centering
\includegraphics[width=0.6\textwidth]{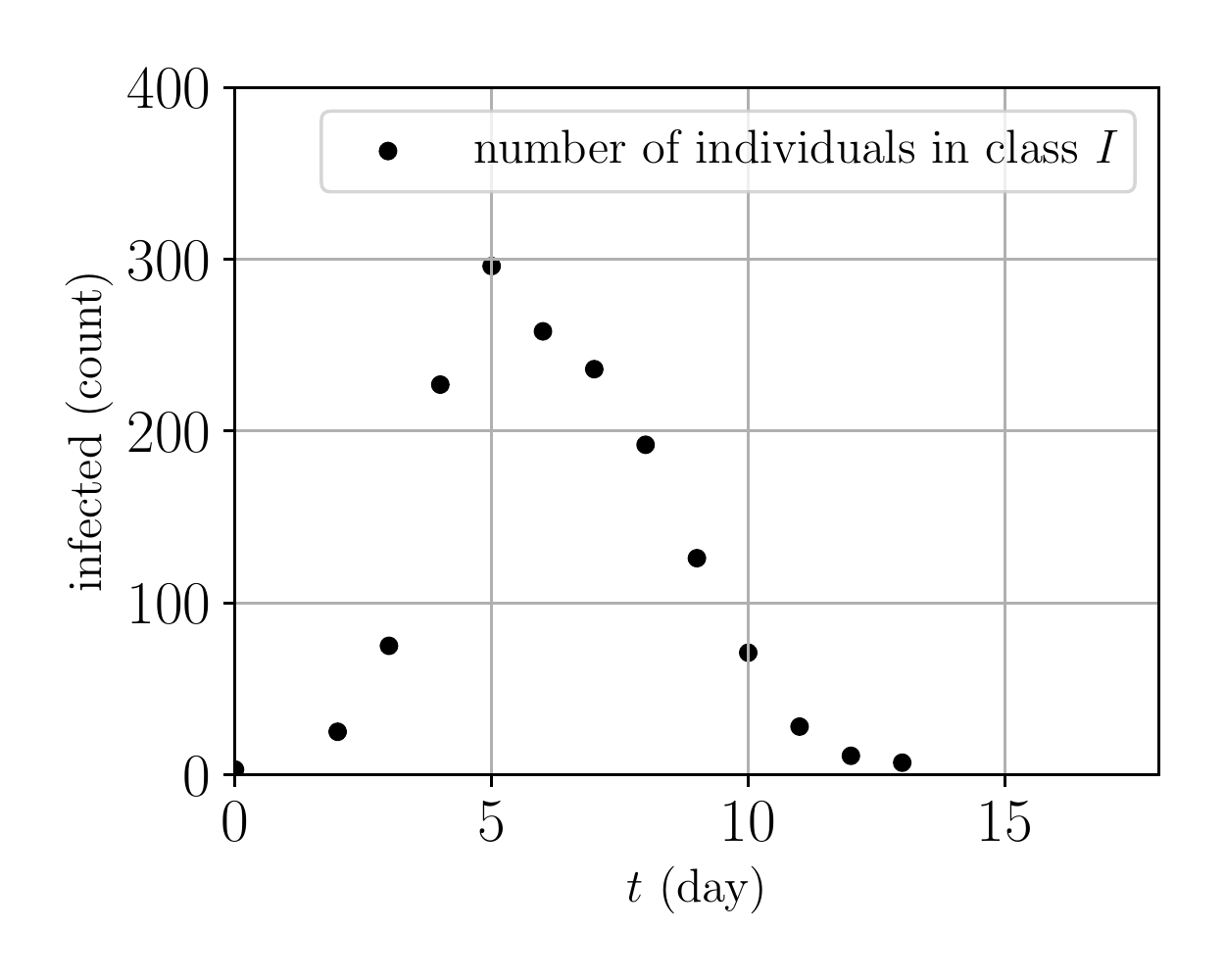}
\caption{Example 3 (SIR): English Boarding School data. The number of infected individuals at the reported times in days from the start of the flu outbreak.}
\label{fig:fludata}
\end{figure}

In this example, we fit
the well-studied Susceptible-Infected-Recovered (SIR) epidemiological model to a classic data set, reproduced in Fig.\,\ref{fig:fludata},
from a flu outbreak at an English Boarding School. \cite{BoardingSchool} The SIR model
has a nearly 100-year history, beginning with the work of Kermack and McKendrick in 1927. \cite{kermack1927contribution}  The simplest version of the model comprises coupled ordinary differential equations (ODEs) and assumes a population of individuals that is closed and well-mixed  in which individuals fall into one of three classes or compartments: susceptible (S), infectious (I), and recovered or removed (R).  The rate of change of susceptible individuals depends on the rate at which new infections arise as a result of contact between infectious and susceptible individuals.  It is typically assumed that the contact rate (number of contacts per unit time) is proportional to the total population size $N$ or constant. The assumption that contacts are proportional to the total population size leads to a mass action term $\beta S I$ for the number of new infections per unit time (called the {\it incidence} in epidemiology). Infectious ($I$) individuals can leave the infectious class through recovery (or death) and progress to the the Recovered (or Removed) class, $R$. It is often assumed that the number of recoveries/removals per unit time is linear in $I$: $\alpha I$.  Under this assumption, the  time spent in $I$ is exponentially distributed with mean $1/\alpha$. The resulting system of ODEs is

\begin{align}
\label{eq:SIR}
    \frac{dS}{dt}&= -\beta SI,\nonumber\\
    \frac{dI}{dt}&= -\alpha I + \beta SI,\nonumber\\
    \frac{dR}{dt}&= \alpha I.
\end{align}

The qualitative dynamics of this model are governed by an epidemiological quantity called the basic reproduction number, $R_0$, which for this SIR model is $R_0 = \beta/\alpha$.  If $R_0>1$, the model exhibits a single outbreak, with $I$ increasing to a peak, then approaching zero as time tends to infinity. Under this scenario, $S$ will decay and approach a positive, constant value, meaning that the epidemic does not infect the whole population.  If $R_0 \le 1$, $I$ will decay and approach zero as time tends to infinity.  This model has played an invaluable role in understanding infectious disease dynamics, informing control strategies, and serving as a building block for more complex compartmental epidemiological modeling frameworks.

The SIR model is often fitted to data by minimizing a  weighted least squares function,
\begin{align}
    F(\theta) = \sum_{n=1}^N w_n  (x_n - I_n)^2 ,
\end{align}
where the data $D = \{ x_1,\cdots,x_N \}$ are the number of infected individuals at the corresponding reporting times $t_1,\cdots,t_N$, $I_n = I(t_n, \theta)$ is the predicted mean infection count at time $t_n$, found by solving Eqs.\,(\ref{eq:SIR}) for the given $\theta$, and $w_n$ are weights. 
Following example 1, we choose a test statistic that is 
transformed and scaled  
\begin{align}
\label{Eq:lambda_SIR}
    \lambda(D, \theta) & = \sqrt{F(\theta) / N} \, / \, 50 ,
\end{align}
so that the statistic is ${\cal O}(1)$.
The weights are set to $w_n = I_n^{-1}$, which we hasten to add should not be taken to imply that the counts $x_n$ are Poisson distributed. On the contrary, the counts $x_n$ are correlated and their fluctuations are super-Poissonian.  

\begin{figure}[h!]
\centering
\includegraphics[width=\textwidth]{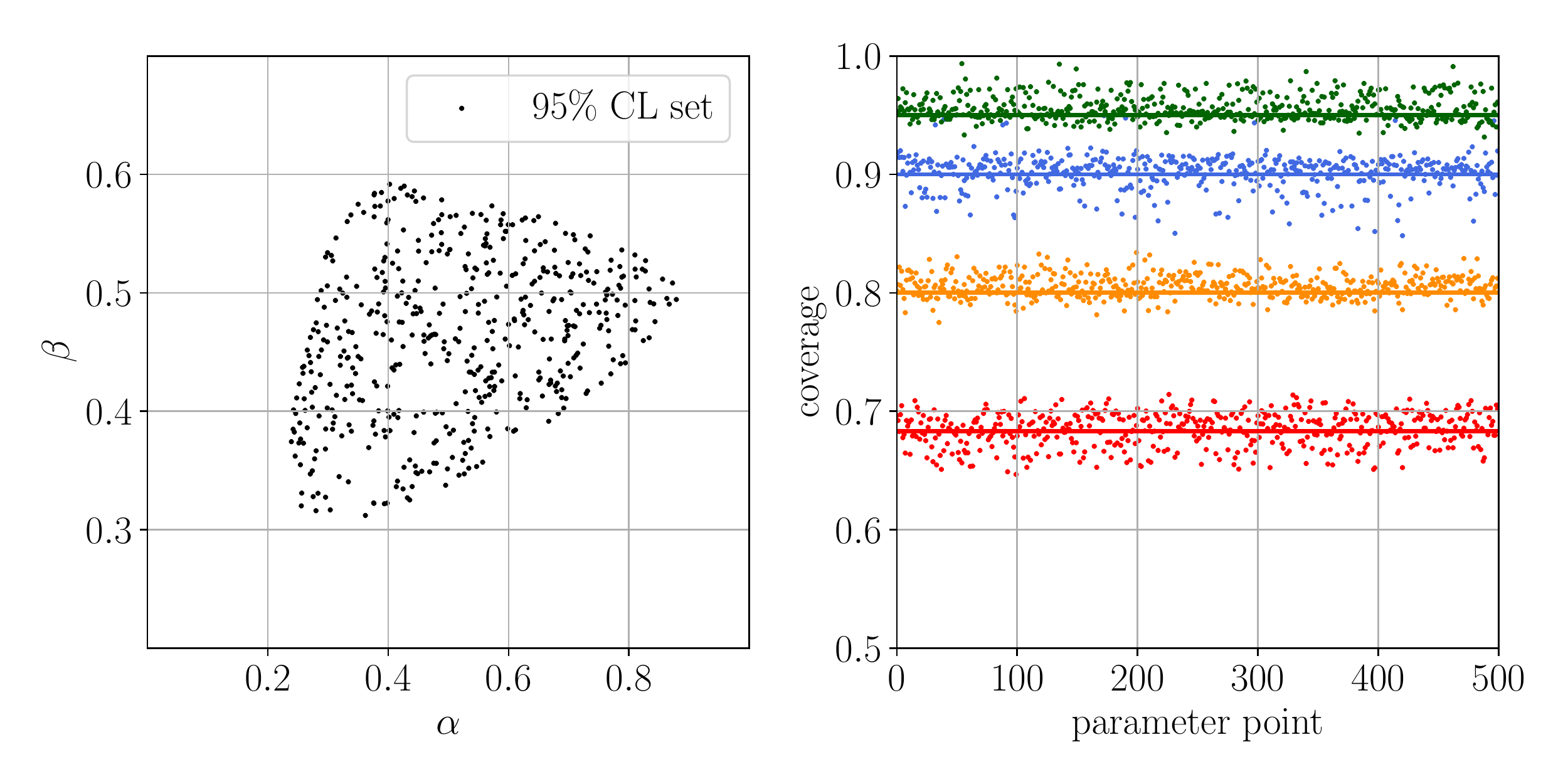}
\caption{Example 3 (SIR): (\emph{left}) 500 points, from the 95\% CL confidence set for the Boarding School data, at which the coverage has been computed in the SIR $\alpha, \beta$ parameter space.(\emph{right}) the coverage probabilities for each of the 500 points in the SIR parameter space.}
\label{fig:SIRcov}
\end{figure}

 We follow a similar training protocol as for example 1, except that the training sample size for this example is 750,000, a fully-connected DNN is used with 6 hidden layers and 25 nodes each, and the best model is found after about 350,000 iterations, that is, after 23 epochs. The confidence sets obtained are shown in 
\begin{figure}[h!]
\centering
\includegraphics[width=0.6\textwidth]{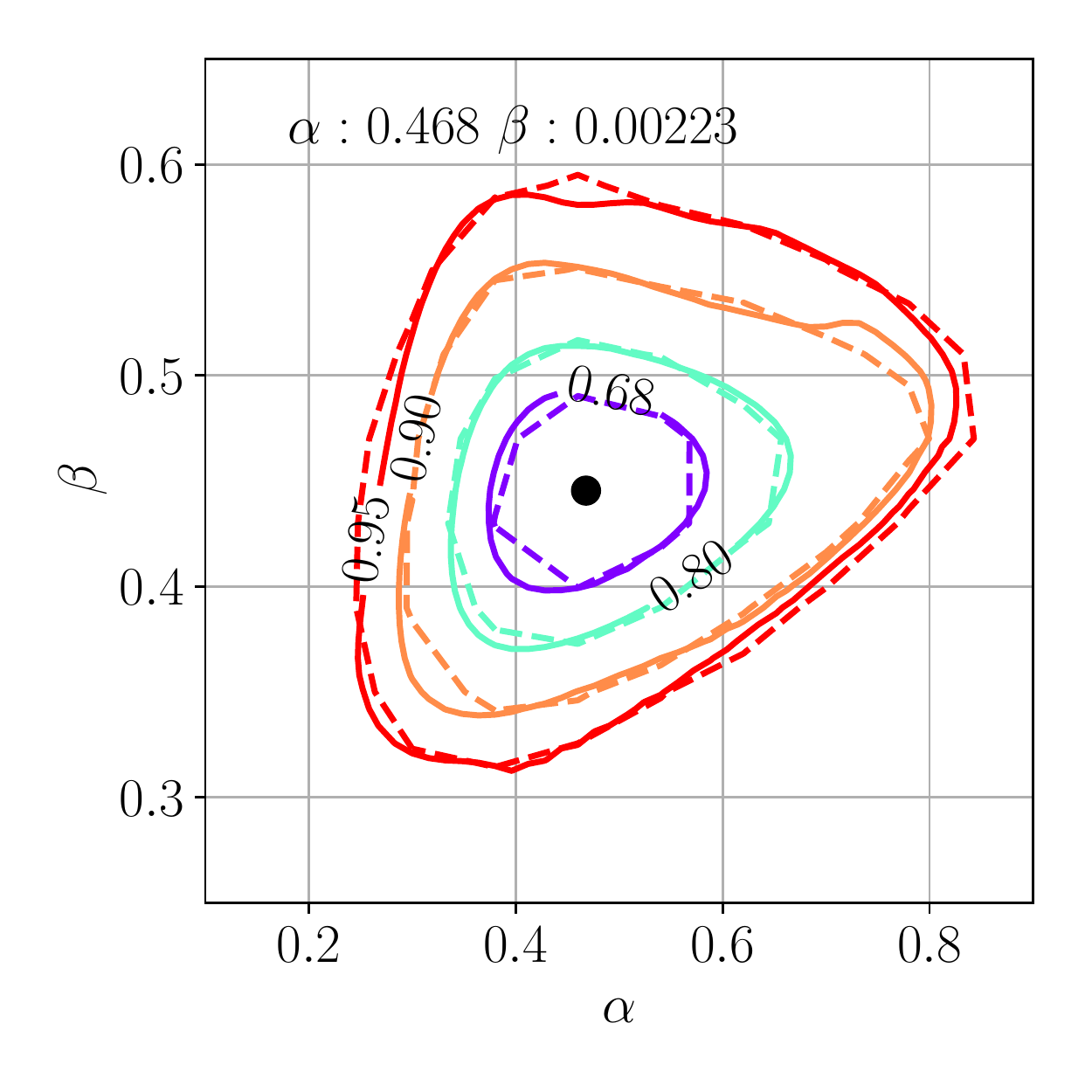}
\caption{Example 3 (SIR): Confidence sets $R(D)$ for $\tau = 0.68, 0.80, 0.90,$ and $0.95$. (dashed lines) Boundaries of 
confidence sets computed with  the histogram-based approximation of the cdf. (solid lines) Boundaries of the confidence sets computed with the DNN-based approximation of the cdf. (\emph{black dot}) Location of the minimum of the DNN-based cdf, which is taken to be the best-fit point.}
\label{fig:SIRsets}
\end{figure}
Fig.\,\ref{fig:SIRsets} and the coverage probabilities are shown in Fig.\,\ref{fig:SIRcov}.
Again, we find that \verb|ALFFI| does an excellent job of creating accurate confidence sets.

\section{Discussion}
\label{sec:discussion}
In the cosmological and epidemiological models all parameters are of interest. But 
in the signal/background-On/Off problem one is typically interested only in the mean signal $\mu$. The mean background, $\nu$, is a nuisance parameter. Unfortunately, constructing confidence intervals for $\mu$ when there are nuisance parameters and when the data are sparse is challenging, though approximate methods exist (see, for example, Ref.\cite{Cowan2011}) including in \verb|LF2I|\cite{dalmasso2023likelihoodfree}.  Given the success of \verb|LF2I| and \verb|ALFFI| in producing reasonably accurate confidence sets, it is of interest to explore whether the reasoning that underlies these approaches can be extended to an algorithm that can yield  provably valid confidence intervals for individual parameters, ideally constructed from the associated multi-parameter confidence set.

One possible approach to construct  confidence intervals from the confidence sets created with \verb|LF2I| and \verb|ALFFI| is to mimic the way that confidence intervals can be constructed from  
 confidence ellipsoids for a multivariate normal density. In the two dimensional example, the objective would be to map a 2-dimensional confidence set to a one-dimensional confidence interval, as is done in the bivariate normal density. For example, in the signal/background problem, one can 
 construct an interval for  $\mu$ as follows: $I(D)  = [\text{min}(\{ \mu_i \} ), \text{max}(\{ \mu_i \} )]$, where $\{ (\mu_i, \nu_i) \}$ are points from the associated confidence set. It should be possible to use an algorithm like \verb|ALFFI| to map from the confidence level of the set to that of the associated interval for the parameter of interest.  If this can be done, then one would be able to determine what confidence level is needed for the confidence set to obtain the desired confidence level for the associated interval. To the best of our knowledge, 
 if
 such a mapping could be devised in the general case it would constitute the first method in which valid multidimensional confidence sets can be mapped to valid one-dimensional confidence intervals without the need for explicit knowledge of the underlying statistical model. Ideas along these lines are under investigation. 

\section{Conclusions}
\label{sec:conclusions}
 The \verb|LF2I| and \verb|ALFFI| approaches are useful when asymptotic results\cite{Algeri:2019lah,Cowan2011} may not be applicable, and are particularly useful when the likelihood function is intractable. The \verb|ALFFI| approach introduced in this paper, which follows \verb|LF2I|, approximates the cumulative distribution function of a test statistic with a neural network in a way that permits a direct check, with the \emph{same} neural network, of the coverage probability of confidence sets at any point in the parameter space of a theoretical model at which sufficient simulated data are available. This provides an \emph{a posteriori} assessment of the accuracy of the confidence sets constructed with \verb|ALFFI|. In addition, by directly binning the point cloud of theoretical model parameters (presumably, in a more sophisticated way than is done in this paper) it is possible to check directly the quality of the neural network approximation. 

\ack
This work was performed in part at the Aspen Center for Physics, which is supported by National Science Foundation grant PHY-1607611. This work is supported in part
by the Department of Energy under Award No. DE-SC0010102 (HP) and supported in part by National Science Foundation grant DMS-2045843 (OP).
We thank Prof. Ann Lee for fruitful discussions at the Aspen Physics Center in summer 2022. 


\newpage

\appendix
\appendixpage
\section*{Appendix: ALFFI algorithm}
\label{alg:ALFFI}
\begin{algorithm}[H]
\caption{Estimate the CDF $\mathbb{C}(\lambda_0 \mid \theta_0) = \mathbb{P}(\lambda < \lambda_0\mid \theta_0)$, given the observed value $\lambda_0$ of a test statistic $\lambda$.} 
\begin{algorithmic}[1]
\Ensure estimated CDF $\widehat{\mathbb{C}}(\lambda_0 \mid\theta)$ for all $\theta=\theta_0 \in \Theta$
\State Set $\mathcal{T'} \gets \emptyset$
\For{$i$ in $\{1,...,B' \}$}
\State Draw parameter $\theta_i \sim \pi_\theta$
\State Simulate $\mathcal{D}_i \leftarrow\left\{X_1, \cdots, X_n\right\}_i \sim F_\theta$
\State Compute test statistic $\lambda_i \gets \lambda(\mathcal{D}_i, \theta_i)$
\State Simulate $\mathcal{D}_i^\prime \leftarrow\left\{X_1, \cdots, X_n\right\}_i^\prime \sim F_\theta$
\State Compute observed test statistic $\lambda_{i}^\prime \gets \lambda(\mathcal{D}_i^\prime, \theta_i)$
\State Compute discrete indicator variable $Z_i \gets \mathbbm{1} (\lambda_i < \lambda_i^\prime)$
\State $\mathcal{T'} \gets \mathcal{T'} \cup \{ ( Z_i, \theta_i, \lambda_{i}^\prime) \}$
\EndFor
\State Use $\mathcal{T'}$ to learn the function $\widehat{\mathbb{C}}(\lambda_0 \mid \theta)$ 
\State \Return $\widehat{\mathbb{C}}(\lambda \mid \theta)$.
\end{algorithmic}
\end{algorithm}


\section*{References}
\bibliographystyle{iopart-num}
\bibliography{AJP_LFI_paper}

\end{document}